\documentclass[letterpaper,11pt]{article}
\usepackage[letterpaper, left=1in, top=1in, right=1in, bottom=1in,nohead,includefoot]{geometry}
\usepackage{color,charter,graphicx,natbib,enumitem,amsmath,amssymb,amsfonts,titlesec,placeins,bm,setspace,caption} 
\usepackage[svgnames,x11names]{xcolor}
\usepackage[T1]{fontenc} 
\usepackage[pdftex]{hyperref}
\hypersetup{
	colorlinks,%
	linkcolor=MidnightBlue,  
	urlcolor=DarkSlateGray,   
	citecolor=RoyalBlue2  
	}
	
\titleformat*{\section}{\normalfont\Large\bfseries\blu}
\titleformat*{\subsection}{\normalfont\large\bfseries\blu}
\titleformat*{\subsubsection}{\normalfont\normalsize\bfseries\blu}
 
\def\blu{\color{RoyalBlue4}}       
\def\bem#1{{\blu\em #1}} 
 
\def\y{\mathbf{y}}\def\balpha{{\bm\alpha}}
\def\seq#1#2{#1{:}#2}
\def\parablu#1{\medskip\noindent{\em\blu#1.}}

\def\defn#1{\medskip\noindent{\em\blu Definition.} #1\medskip}
\def\bi{\begin{itemize}[noitemsep,topsep=3pt,itemsep=-3pt]}
\def\ei{\end{itemize}}
\def\ci{\perp\!\!\!\perp} 
\def\eqn#1{eqn.~(\ref{eq:#1})}\def\Eqn#1{Eqn.~(\ref{eq:#1})}

\def\beq#1{\begin{equation}\label{eq:#1}}\def\eeq{\end{equation}}
\def\tE{\textrm{E}}   \def\tN{\textrm{N}}   \def\tV{\textrm{V}}\def\tP{\textrm{P}}
\def\e{\textrm{e}}
\def\tPr{\textrm{Pr}}  
\def\cH{\mathcal{H}}\def\cS{\mathcal{S}}  
\def\scn{\cS}\def\scnj{\scn_j}
\def\emr{\pi_{pf}}\def\emrpf{\emr}\def\emrfp{\pi_{fp}}
\def\emra{\pi_{pf}(\balpha)} 
\def\pa{\rho(\balpha)}
\def\py{p(\y)}\def\pjy{p_j(\y)}\def\fya{f(\y|\balpha)}\def\ry{r(\y)}
\def\alh{\widehat{\balpha}}\def\als{\balpha^*}\def\fy{f(\y)}\def\ky{k(\y)}\def\logpostalpha{\lambda(\balpha)} 
\def\kpf{\kappa_{pf}} \def\kfp{\kappa_{fp}} 

\def\a{\mathbf{a}}\def\tDir{\textrm{Dir}}\def\p{\mathbf{p}}\def\h{\mathbf{h}}\def\H{\mathbf{H}}
\def\bone{\mathbf{1}}
      \def\klpf{\kpf} \def\klfp{\kfp}  
      \def\esspf{\epsilon_{pf}} 
\def\tDir{\textrm{Dir}}
\def\safehaven{backstop }\def\safehavennospace{backstop} 
 
0
\graphicspath{ {./figures/} }

\newcommand{\blind}0 

\begin{document}

\begin{center} 

{\blu\bf\LARGE Predictive Concordance for Parameter Optimisation\\ and Mixture Synthesis}

\if0\blind
	{ \bigskip
		{\large  
			    Tobias Adrian,\footnote{{\blu Tobias Adrian}, Director of the Monetary and Capital Markets Department, International Monetary Fund 
                							    \\ \indent\indent\indent  700 19th Street NW, Washington, DC 20431, U.S.A.  
		  							     \\ \indent\indent\indent \href{mailto:tadrian@imf.org}{tadrian@imf.org}}
                Domenico Giannone,\footnote{{\blu Domenico Giannone}, Bloomberg Distinguished Professor of Economics and Statistics, Johns Hopkins University 
                							    \\ \indent\indent\indent  Wyman Park Building 5th Floor, 3100 Wyman Park Drive, Baltimore, MD 21211, U.S.A.  
		  							     \\ \indent\indent\indent \href{mailto:domenico.giannone@jhu.edu}{domenico.giannone@jhu.edu}}		     
                Matteo Luciani\footnote{{\blu Matteo Luciani}, Principal Economist, Board of Governors of the Federal Reserve System
	         							   \\ \indent\indent\indent 20th Street and Constitution Avenue NW, Washington, DC 20551, U.S.A.
	        							    \\ \indent\indent\indent \href{mailto:matteo.luciani@frb.gov}{matteo.luciani@frb.gov}} 
			   and Mike West\footnote{{\blu Mike West}  The Arts \& Sciences Distinguished Professor Emeritus of Statistics \& Decision Sciences   
	   	  							 \\ \indent\indent\indent Duke University, Durham, NC 27708, U.S.A. 
		 							  \\ \indent\indent\indent \href{mailto:mike.west@duke.edu}{mike.west@duke.edu}
		 							  \\ \indent\indent\indent  {\em Corresponding author}}
		}
	} \fi  

\bigskip\bigskip

{\blu\bf \Large Abstract} \end{center} 
We discuss probabilistic measures of concordance between two probability distributions based on the expected misclassification rate (EMR). 
The focus is on comparing a given reference distribution with other distributions in a parametrised class, and optimising concordance by identifying 
parameter values maximising EMR or a regularised variant.  EMR is a practical and decision-theoretically meaningful measure,
and its optimisation has direct interpretation as a Bayesian decision analysis with a bounded utility function. 
We explore theoretical properties of EMR, discuss relationships with other measures including K\"ullback-Leibler divergence,  and recognise 
that its optimisation has a synthetic Bayesian emulation interpretation that aids understanding and specification of regularisation penalties. A main 
area of methodology is in mixture synthesis where the parametrised family is a discrete mixture of given distributions. A detailed example comes from scenario forecasting in macroeconomic policy settings, a key applied area motivating the new methodology.  Theoretical developments underlie efficient numerical optimisation and analysis is easily implemented using direct Monte Carlo simulation. 

{\em What-if??} forecasting

\setstretch{1.10} 
\thispagestyle{empty}
\if0\blind
	{   \renewcommand{\thefootnote}{ } 
            \footnotetext{\\[3pt] \noindent {\bf Disclaimer}: The views expressed in this paper are those of the authors and do not necessarily reflect the views and policies of the Board of Governors, the Federal Reserve System, or the International Monetary Fund, its Management, or its Executive Directors.}
            \renewcommand{\thefootnote}{\arabic{footnote}}
            \setcounter{footnote}{0}
	} \fi  
 
\newpage

\section{Introduction}

This paper concerns the evaluation and quantification of concordance of a parametrised family of distributions with a defined reference distribution.  Such questions arise in several areas of statistical model emulation,  evaluation and assessment of assumptions underlying predictions.   The focus is on assessing 
the parameters of a defined class of parametrised distributions to best match a distribution that arises from an assumed data generating theoretical, physical or empirical statistical model. The latter distribution is adopted as a representation of  so-called ground truth, a benchmark to measure 
comparative relevance of distributions in the parametric family.  

The setting is that faced by a decision maker (DM) interested in a future vector outcome $\y$.   To represent their information and uncertainty about $\y,$ 
DM adopts a family of predictive distributions with density functions $\fya$ known up to a parameter vector $\balpha$.  DM is then presented with a ground truth density $p(\y)$-- designated the {\em reference density}--  and is interested in how the chosen parametric family can best approximate the 
reference in predicting the (future) outcome $\y.$   This question for DM translates to that of assessing parameters $\balpha$ so that the resulting $\fya$ best concords with the reference.  

Of prime interest here are contexts in which DM's predictions define a discrete mixture of fully specified predictive density functions (p.d.f.s), the latter corresponding to predictions from each of a set of models or assumed scenarios. Here the parameter  
 $\balpha$ is the vector of mixture probabilities that relatively weight predictions from each of the models.  One main applied setting that originally motivated the current developments is in scenario forecasting in economics~\citep{Bernanke2024}, where DM is indeed faced with model uncertainty: a given set of predictive p.d.f.s  represent forecasts based on each of a selected set of contextual assumptions in one or more economic models. These represent a set of  scenarios chosen by DM to be assessed relative to an empirically well-founded representation of uncertainty about the variables of interest, the reference distribution from a data-based statistical model.  Optimising the mixture probabilities $\balpha$ to best match the reference then  addresses interests in 
evaluating the comparative relevance of the scenarios. 

Related contexts arise in so-called {\em What-if?} forecasting in business and commercial applications, and in areas of financial risk and portfolio analysis where sets of new and evolving models are routinely evaluated against an ongoing standard or reference model.  In these and the macroeconomic forecasting settings, it is typical that the reference distribution is represented in terms of simulation samples, so that analysis will be based on some form of Monte Carlo analysis. 

This is not a traditional parameter inference setting, i.e., a context where observed data is used to inform on parameters in the usual (Bayesian or non-Bayesian) sense.  There is no observed data here; DM's information is that $p(\y)$ represents the outcome $\y$ that is to be observed but is not currently available.  The questions are then those of  comparing $\fya$ with $p(\y)$ across ranges of possible values of $\balpha.$  This raises the main question of how to measure concordance of distributions, i.e., of choosing a utility function to characterise and quantify closeness when comparing densities. Our preferred measure comes from traditional statistical classification as discussed in following sections, with connections to  other related measures. 

\section{Predictive Concordance via Misclassification Probabilities\label{sec:ConcordanceEMR}} 
 
For notational simplicity and clarity in this Section, we use $\fy$ in place of the parametrised p.d.f. $\fya$ throughout, and assume 
that $\py$ and $\fy$ have the same support. 
 
\subsection{Expected Misclassification Rate: Foundation and Definition\label{sec:EMR}} 

Our preferred measure of predictive concordance is based on statistical misclassification probabilities that are natural and interpretable, and that 
have appealing properties. They are also somewhat underutilised in statistics to date, so the development here is of broader interest.    

Suppose that a random draw $\y$ is made from either $\fy$ or $\py$ with equal probabilities. It is not disclosed which distribution generates the outcome $\y.$ Write $\cH_p$ for the hypothesis that $\y\sim\py,$ and $\cH_f$ for the hypothesis that $\y\sim\fy.$ 
Since the choice is made with $\tPr(\cH_p)=\tPr(\cH_f)=0.5,$ the resulting
posterior probabilities conditional on the observed $\y$ are   
$ \tP(\cH_p|\y) = \py/\{\py+\fy\}$ and $\tPr(\cH_f|\y)=1-\tP(\cH_p|\y).$ 

Now assume that $\y$ is actually a draw from $\py$, i.e., condition on $\cH_p$.   
Before learning $\y,$ the expected posterior probability on $\cH_f$ is then 
\beq{EMR} 
\emr \equiv \tE[\tP(\cH_f|\y) |\cH_p] = \int_\y  \tP(\cH_f|\y) \py d\y =  \int_\y  \frac{\fy\py}{\{\fy+\py\}}d\y.
\eeq
By symmetry,  if $\y$ is actually from $\cH_f,$ the 
expected posterior probability $\emrfp= \tE[\tP(\cH_p|\y) |\cH_f]$ is obviously the same, $\emrfp = \emr$.
 
\defn{\Eqn{EMR} is the expected misclassification rate (EMR)  in discriminating  $\fy$ and $\py$.}

EMR is an inherent measure of so-called confusion and interpretable on the absolute probability scale. 
Higher values indicate that  it is difficult to discriminate  $\fy$ from $\py$-- indicating that draws from $\fy$ are more likely to be misclassified as coming from $\py$--  and vice-versa.   This is a natural, interpretable measure of concordance-- or discordance-- of the two distributions.  
In traditional classification in statistics and machine learning, the optimal Bayesian classifier judges $\y$ as coming from $f(\y)$ with probability $\tP(\cH_f|\y).$ Averaging across $\y\sim p(\cdot),$ and using standard terminology,  
$1-\pi_{pf}$ is then both the population  sensitivity 
and (due to the comparison of just two distributions and the implied symmetry) 
the population  specificity 
of the optimal Bayesian classifier. It follows that $1-\pi_{pf}$ is the  traditional overall accuracy of the test comparing $f(\cdot)$ and $p(\cdot)$, and so EMR $\pi_{pf} =1-\textrm{accuracy}$ is the traditional error rate. Increasing EMR indicates decreased discrimination of $f(\cdot)$ from $p(\cdot)$. Judging $f(\cdot)$ to be
close to $p(\cdot)$ at higher values of $\pi_{pf}$ is thus theoretically fundamental and practically interpretable.

\subsection{EMR Properties, Connections and Examples\label{sec:EMRaspects}} 
 
It is immediate that $\emr\le 0.5$ with equality only when $f(\cdot)\equiv p(\cdot),$ defining the absolute scale for assessment of concordance.  To prove this, simply note that  
$\emr = \tE[r(\y)/\{1+r(\y)\} |\cH_p]$ where $r(\y)=\fy/\py$ with $\tE[r(\y) |\cH_p]=1.$ 
Now, $r/(1+r)$ is concave on $r>0$ so   that  
$\emr \le \tE[r(\y)|\cH_p]/\{1+\tE[r(\y)|\cH_p]\}=1/2.$  
The upper bound is achieved when $f(\y)\equiv p(\y)$.  Correspondingly,  $1-\emr$ is a measure of difference between $\fy$ and $\py$, also on the interval $(0,1/2]$ with higher values indicative of greater difference. 

EMR is one of a number of measures of concordance between distributions. Traditional measures such as  K\"ullback-Leibler (KL),  total variation (TV) and mean absolute p.d.f. differences are close cousins. A key distinction is that EMR is wholly focused on subjective perception of difference, in terms of the probabilistic ability to discriminate. Other measures are less transparent and interpretable in terms of their scales-- especially when $\y$ is in more than one dimension. EMR also relatively down-weights fluctuations of relative p.d.f. values in the tails, again with a  main focus on the aggregate probabilistic  determination of differences in expected outcomes. KL divergence-- arguably the most popular and theoretically supported measures of distributional concordance-- is of main interest. The KL divergence {\em of} $f(\cdot)$ {\em from} $p(\cdot)$ is $\klpf \equiv  \tE[k(\y) |\cH_p]$   where
$\ky=-\log\{r(\y)\}$ with $r(\y)=\fy/\py$ as defined earlier.   This compares with $\emr = \tE[ \ry/\{1+\ry\} |\cH_p]$.    
A related measure is the
theoretical {\em effective sample size} (ESS) from Monte Carlo (MC) importance sampling (IS). In using $\py$ as an 
importance sampling {\em proposal distribution} for MC evaluations of expectations under {\em target distribution} $\fy,$ the ESS is $\esspf = \{ \tE[r(\y)^2|\cH_p]\}^{-1}$ on the $(0,1]$ scale, with perfect agreement at $\esspf=1.$

\parablu{Odds, Probabilities and Existence} KL evaluates expected odds ratios in comparing $\fy$ to $\py$, whereas EMR evaluates expected probabilities. 
An under-regarded point about KL is that there are important practical contexts where KL is undefined as a result of the critical dependence on tails of the log p.d.f.s involved.  Simple examples arise when  $\fy$ is Gaussian and $\py$ does not have finite variance.  In contrast, $\emrpf\in (0,0.5]$ always.  This is, in part, a strong reminder of the relevance of  {\em bounded} utility functions in decision analysis. While generally to be stressed,  this arises here with  respect to EMR as an expected utility function in a decision problem where $\fy$ is to be chosen to approximate $\py$ with the bounded utility function $f(\y)/\{f(\y)+p(\y)\}$ when $\y\sim p(\y).$

\parablu{Approximations and Bounds}  Define $\eta(k) = 1/\{1+\exp(k)\}$ for real $k$ and consider approximations of $\eta(k)$ around $k=0$. 
The first two derivatives of $\eta(k)$ are 
 $\dot\eta(k) = -\eta(k)\{1-\eta(k)\}$ and $ \ddot\eta(k) = \eta(k)\{1-\eta(k)\}\{1-2\eta(k)\}$. These  underlie the following results.  
When $k$ stands for  $k(\y)$ under $\cH_p,$   the first-order (delta) approximation yields $\emr\approx\eta(\klpf)$.  Hence any comparison of $\fy$ and $\py$ will yield the same results under concordance based on KL as with EMR to this level of approximation. This is relevant when $\fy$ and $\py$ are close in terms of either mmeasure. 
The second-order approximation refines the above to $\emr\approx\eta(\klpf) + s_p\ddot\eta(\klpf)/2$, adding the correction term in which 
$s_p = \tV[k(\y)|\cH_p]$, the variance of $k(\y)$ under $\cH_p.$  Since the correction terms is positive unless $\fy=\py$ everywhere, this implies that 
$\emr \ge \eta(\klpf)$  to this degree of approximation;  i.e., the inverse logistic transform of KL is an approximate lower bound on EMR. 

\newpage

\begin{figure}[htbp!]
\centering
\includegraphics[width=3.0in]{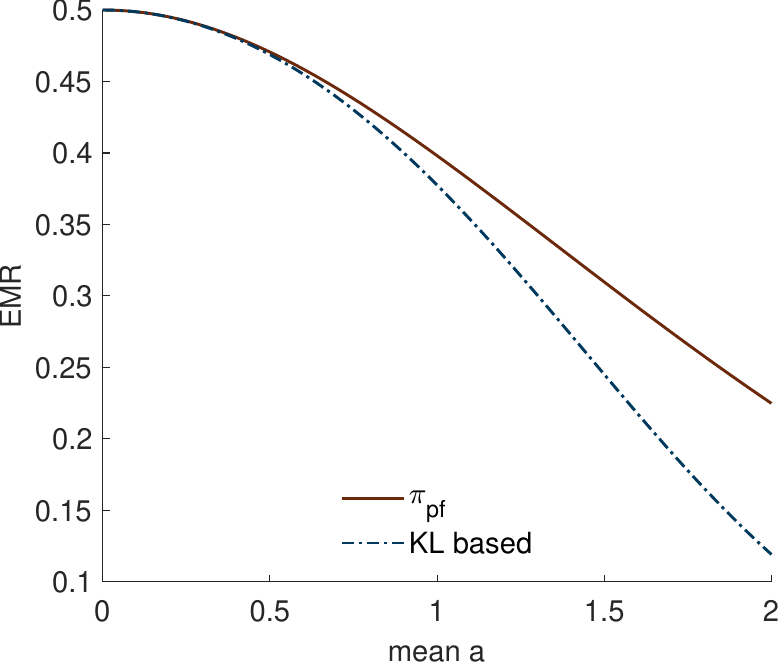}  
\includegraphics[width=3.0in]{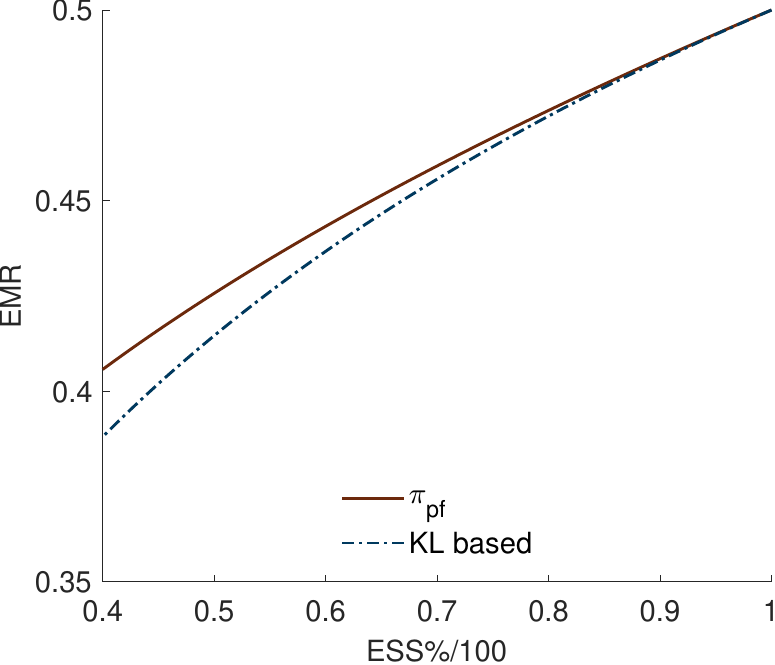}  
\caption{Predictive Concordance Example: EMR, ESS and KL-based lower bound when $p(y)=\tN(0,1)$ and $f(y)=\tN(a,1)$ for a range of values of $a$.
 \label{fig:piessnormalEG}}
\end{figure}

\parablu{A Simple Example} 
Take $\y=y$ to be scalar with $p(y) =\tN(0,1),$  standard normal, and $f(y)=\tN(a,1)$ for some mean 
$a\ge 0.$   Here $\klpf=a^2/2$ and  $\esspf = \exp(-a^2) = \exp(-\klpf^2).$   
Figure~\ref{fig:piessnormalEG} compares $\emr$, the lower bound $\eta(\klpf)$ and $\esspf$ on a range of values of $a\ge 0$.  
For $a\le 1,$  $\esspf\ge 0.4,$  while $ \emr\ge 0.4$  and is roughly linear in $\esspf$ up to its maximum of 0.5. For practical purposes and
extrapolating from this interpretable example-- also supported by other empirical examples--  $\emr\ge 0.4$ or so is expected unless $f(\cdot)$ and $p(\cdot)$ are quite substantially discordant.
Also, 
the first-order approximation  $\pi_{pf} \approx 1/\{1+\exp(\kpf)\}$ is very accurate when $\emr> 0.4$. 

\FloatBarrier

More generally, the link of EMR to KL is further illuminated via the general approximation
 $\eta(k) \approx (2-k)/4$ for small $k.$   This is both the 1st- and 2nd-order Taylor series approximation of $\eta(k)=1/\{1+\exp(k)\}$  at $k=0.$ 
 It can be shown that   this is an exact lower bound on $\eta(k)$  for $k\ge 0$ and an exact upper bound for $k\le 0.$
This approximation is very accurate over $|k|\le 0.5$ where $\eta(k)\ge 0.38$; the absolute error is less than 
$0.68\%$ on $|k|\le 0.5$.  Hence, if $\y\sim \py$ and the implied distribution of $k(\y)$  heavily favours   $|k(\y)|\le 0.5,$
 then $\emr \approx \{2-\kpf\}/4.$  On this basis, maximising $\emr$ is 
  approximately minimises KL divergence.

\parablu{Classes of Theoretical Examples} The lower bound insights above are not general; there are obvious examples in which it does not arise.  That said, there are practical settings and examples when it is theoretically implied and from which useful insights are gleaned.   The simple normal example above is one very special case. A more general class of examples is that of scale mixtures of normals. 
Under $\cH_p,$ denote by $g(k)$ the p.d.f. of $k(\y) = \log\{\py/\fy\}$ and assume this has finite mean $m$; the mean will, of course, be non-negative.
Then, if  $g(k)$ is continuous, symmetric and unimodal at $m,$ it can be shown that $\emr\ge \eta(\klpf)$ is theoretically implied. Details are in 
Appendix~A.    As elaborated in that appendix, it seems likely that the theory extends to
 more general cases when $g(k)$ is not symmetric, though this is an aside and beyond the main interests of the current paper. 
Importantly, the above discussion does not extend at all to cases-- including many practical cases-- when the
KL divergence does not exist and/or when the distribution $k(\y)$ is less regular.

\newpage

\parablu{Symmetry Considerations}  KL is directional, with $\klpf$ and $\klfp$ typically different measures. 
In contrast, the symmetry of EMR in $f(\cdot)$ and $p(\cdot)$ implies that the same results hold with the two densities exchanged.  
Then, in cases of the  lower bound noted  above, the symmetry of EMR leads to the implied 
to $\pi_{pf} \ge 1/\{1+\exp(\kappa)\}$ where $\kappa= \min\{ \klpf,  \klfp \}.$  Under circumstances when the lower bound applies-- and in terms of general insights-- this directly addresses the
question of directional definition of KL; the relevant symmetrisation of KL is $\kappa$  (rather than an arithmetic or other average of the two directional KL divergences).

\section{Parametrised EMR: Optimisation and Interpretations} 

\subsection{Optimisation Setting} 

Return now to the setting where $\fy \equiv \fya$ is a parametrised family of p.d.f.s and DM is evaluating parameters $\balpha$ relative to the implied predictive concordance with the given reference p.d.f. $\py.$   Recognising $\balpha$ in notation, \eqn{EMR} is now
\beq{EMRa} 
\emra =  \int_\y  \frac{\fya\py}{\{\fya+\py\}}d\y.
\eeq
DM aims to explore values of $\balpha$ such that  $\fya$ is close to $p(\y)$ in terms of high values of $\emra.$ 
The setting is general in terms of possible classes of parametrised p.d.f.s $\fya.$   

DM regards $\fya/{\{\fya+\py\}}$ as their utility function in choosing $\balpha$ in the face of uncertainty about $\y$ represented by $p(\y).$  
This is a natural and (desirably) bounded utility. The expected utility to be optimised is \eqn{EMRa}, with maximiser denoted by 
$\alh = \textrm{argmax}_{\balpha}\emra.$     DM will also be interested in aspects of the implied distribution of utilities, i.e., the distribution of 
$f(\y|\alh)/\{f(\y|\alh)+\py\}$ under $\cH_p$ (see~\citealp{West2024constrainedforecasting}, for example, for examples of the importance of exploring predictive distributions of optimised utilities, as well as simply operating with the expected utility maximising decision). 

\subsection{Synthetic Likelihood Interpretation\label{sec:EMRlike}}
 
Suppose that reality $\y$ is generated from the reference $p(\y)$ and 
consider an hypothetical binary outcome $z$ generated from the Bernoulli distribution 
with success probability $\tPr(z=1|\y,\balpha) = f(\y|\balpha)/\{p(\y)+f(\y|\balpha)\}.$ Then
$$\tPr(z=1,\y|\balpha) = \tPr(z=1|\y,\balpha) p(\y|\balpha) = \tPr(z=1|\y,\balpha)p(\y) 
= \fya\py/\{\fya+\py\}.$$ 
Now suppose DM observes $z=1$ but not $\y$; EMR emerges via expectations over the \lq\lq missing data" $\y,$  viz., $\tPr(z=1|\balpha) = \emr(\balpha).$   Thus, $\emr(\balpha)$ is in fact a likelihood function for the {\em parameter} $\balpha$ based on an hypothetical observation $z=1$ that classifies a random draw from $p(\y)$ as coming from $\fy$ under a 50:50 prior.   
It follows that EMR-maximiser $\alh$ is a maximum likelihood estimate (MLE) in this purely synthetic likelihood setting.

\subsection{Synthetic Posterior and Regularisation\label{sec:EMRpost}}
 
The above discussion promotes a Bayesian emulation perspective with synthetic likelihood modified by a synthetic prior, and optimisation seeks out the posterior mode for $\balpha$ rather than the MLE.  In this optimisation setting, the prior can be regarded as simply defining a penalty term that stochastically constrains the likelihood-only analysis.  This turns out to be particularly interesting in the first main setting of applied interest, that in which $\fya$ is a discrete mixture of given p.d.f.s. developed in Section~\ref{sec:mixturesEMR} below.   Here, at a general level, if DM adopts an
 hypothetical prior p.d.f. $\pa$ then-- on the log scale-- the objective function to maximise with respect to $\balpha$ is 
\beq{Logpostalpha} 
\logpostalpha =  \log\{\emr(\balpha)\} + \log\{ \pa \}. 
\eeq
The posterior mode $\als$ modifies the synthetic MLE $\alh$ based on the regularisation penalty $ \log\{ \pa \}. $
 
With respect to other measures of concordance, such as KL and TV already noted, a key point relates to optimisation regularisation interests.  The theoretical   interpretation of EMR as a synthetic likelihood naturally engenders this Bayesian interpretation in which the penalty function relates to a synthetic prior.  As a result, its choice and calibration is directly interpretable.  This adds to interests in EMR and contrasts with the lack of intuitive, interpretable scales for other measures. 
 
There are direct connections with the statistical literature on optimisation using purely synthetic, mathematically equivalent formulations in terms of 
a Bayesian analysis where a defined objective function can be recapitulated as a (log) posterior density for parameters under a purely hypothetical statistical model. Early developments of \cite{mueller1999} on this concept have led to ranges of applications in areas including biostatistical design and portfolio analysis~\citep[e.g.][]{EkinPolsonSoyer2014,IrieWest2018portfoliosBA}. The context here is similar in that the EMR objective function has the synthetic likelihood/posterior interpretation. 
This literature connection highlights the point that this synthetic posterior interpretation opens access to numerical methods-- including Monte Carlo Markov chain, for example-- to explore the $\lambda(\balpha)$ surface, perhaps especially around the maximising value for sensitivity analyses. That said, in our main setting of mixtures below, we show that the optimisation is provably  convex so standard optimisation algorithms apply directly in that setting.

\section{Predictive Mixture Synthesis\label{sec:mixturesEMR}}  

\subsection{Mixture Setting} 

Suppose that DM's predictions are based on a discrete mixture of fully specified predictive p.d.f.s
\begin{equation}\label{eq:scenariomixturef} 
\fya = \sum_{j=\seq 0J} \alpha_j p_j(\y)
\end{equation}
where the $p_j(\y)$ are given and $\balpha = (\alpha_0,\ldots,\alpha_J)'$ is a probability vector. This is a very general context of DM having a set of $J+1$ models with model uncertainty; the $\alpha_j$ define the model probabilities that relatively weight predictions from each of the models.  One main applied setting that originally motivated the current developments is in scenario forecasting in economics and business, where the mixands $p_j(\y)$ are forecast p.d.f.s based on each of a selected set of models and/or contextual assumptions in one model.   These represent a set of {\em scenarios} chosen by DM as of relevance in understanding concordance with the empirically supported reference distribution.  For this reason we refer to the $\alpha_j$ as {\em mixture weights} and/or {\em scenario probabilities} interchangeably. 


\subsection{EMR Optimisation in Mixtures}

Evaluating the EMR-maximiser $\alh$ is a probability simplex constrained optimisation problem. This is provably convex with a unique solution. We present the details and prove this in Appendix~B.    

The solution $\alh$ will typically be a \bem{sparse mixture} of scenarios, with some zeros in $\alh.$  This follows
from general results of optimisation of convex functions over the probability simplex~\citep[e.g.][]{BoydVandenberghe2004}.
For some integer $k\in \{0:J\}$ a subset of $k$ of the $\widehat\alpha_j$ can be zero. There are cases when $k=0$ but $k>0$-- defining a sparse optimising vector-- is more usual, especially with larger $J$ and diversity among the $p_j(\y).$ 
This relates to general features of optimisation over the simplex; simplex constraints operate to shrink weights to the boundaries, effectively as  $\ell_1$ shrinkage for sparsity~\cite[e.g.][]{brodie2009sparse}. Sparsity aids in scalability to larger numbers $J$ of mixands.

However, sparsity in $\alh$ is unstable since it is not a genuine feature but  is induced by the implicit prior $\ell_1$ penalty; its values are typically very sensitive to small changes in the input p.d.f.s in the mixture and the reference. This pathology of sparsity inducing penalties was identified and documented in the context of forecasting by \citet{IllusionOfSparsity}. In the current setting, take an example with two very similar scenario p.d.f.s; one of these scenarios will have a zero value in $\alh,$ the other non-zero.  Then, a very small change in either of the p.d.f.s-- or of the reference p.d.f.-- will flip the zero/non-zero pattern.  At each of these extremes-- and for ranges of the $\alpha_j$ on these two scenarios 
bridging the extremes-- the resulting scenario mixture $f(\y|\alh)$ will be almost unchanged. This  sensitivity is undesirable; it is desirable to have similar probabilities on the two scenarios. 

The key point is that $\emra$  function can and often does have modes at the simplex boundaries. This can be addressed by imposing additional constraints or, more foundationally, with a minimally informative regularisation penalty,  i.e., a synthetic prior over $\balpha$ following Section~\ref{sec:EMRpost}.

\subsection{Regularised EMR Optimisation\label{sec:mixturesEMRregularised} }
  
The natural class of synthetic priors are Dirichlet, $\balpha\sim \tDir(\a)$ having p.d.f. 
$\pa \propto \prod_{j=\seq 0J} \alpha_j^{a_j-1}$ over the simplex. Here $a_j>0$ for all $j$
and, with precision $a=\sum_{j=\seq 0J}a_j,$ the means are $a_j/a$ and 
prior joint mode has elements $\max\{0, (a_j-1)/(a-J-1)\}.$ A prior with each $a_j=1+\epsilon$ for small $\epsilon>0$ is \lq\lq minimally informative" subject to the joint prior mode being positive on each scenario.   Modifications to $a_j=1+\epsilon_j$ to differentially favour scenarios {\em a priori} are obviously of interest,  but for this paper the symmetric prior is adopted. For given $\epsilon,$  the prior joint mode and mean are then each $\bone/(J+1),$  i.e., favouring a uniform set of scenario probabilities though with high uncertainty when $\epsilon$ is very small.  Under this prior $\balpha\sim \tDir(\bone(1+\epsilon)),$  the 
regularised objective function of \eqn{Logpostalpha} specialises to 
\beq{LogpostalphaDirprior} 
\logpostalpha =  \log\{\emr(\balpha)\} + \epsilon\sum_{j=\seq 0J} \log(\alpha_j), \quad\textrm{subject to } \alpha_j>0 \ (j=\seq 0J) \quad\textrm{and  } \sum_{j=\seq 0J}\alpha_j=1,
\eeq 
The synthetic prior-based penalty explicitly acts to move from the boundary zero MLE values in $\alh$ to small but non-zero values in $\als$; 
this leads to more stable and robust results and addresses the issues discussed in the previous section.   Further,  the fact that Dirichlet distributions are strongly unimodal means that \eqn{LogpostalphaDirprior} is strictly unimodal as was proven above in the case $\epsilon=0.$  Hence standard and efficient convex optimisation methods apply to evaluate $\als$ as well as $\alh.$ 
 
Analysis requires choice of a (small) value of 
the regularising hyper-parameter $\epsilon.$ Details of calibration in Appendix~C underlie the default recommendation $\epsilon = c/(J+1),$ where $c=0.005.$ The value of $c$ can be modified up/down with minimal impact, while the scaling with number of scenarios is important in more heavily penalising the MLE-based analysis in higher dimensions.  
 
It is also of interest to consider analyses with additional constraints on $\balpha.$ This could, for example, require a (small) lower bound on each $\alpha_j,$ 
and/or constrain a partial ordering of the $\alpha_j.$ Such constraints simply modify the synthetic Dirichlet prior by the indicator of the constraints and 
are again easily incorporated in standard  optimisation algorithms. 

\subsection{EMR Evaluation using Monte Carlo Integration\label{sec:compute}} 

EMR optimisation is  implemented with  standard algorithms for constrained convex optimisation.  Implementation requires the evaluation of $\emra$ in \eqn{EMRa} at candidate values of $\balpha$. {\em Prima facie}, this relies on evaluating the p.d.f.s $\py$ and each $\pjy,$  and then performing the integration. Analytic approximations to the integral may be explored;  useful approximations that relate to measures of discriminatory information in classification using mixtures~\citep[e.g.][]{LinChanWest2015Biostatistics} can be helpful. 
In practice, however,  forecasts from the designated reference distribution will often-- if not typically-- be available in terms of Monte Carlo samples. Hence $\emra$  evaluations are generally addressed using  Monte Carlo integration.
A random sample $\y^i$, $(i=\seq 1n)$, drawn from $\py$ underlies direct Monte Carlo evaluation of EMR in \eqn{EMRa} at any given $\balpha.$  Note also that mixture component-specific EMR values in which $\fya$ is replaced by $p_j(\y)$ can also be directly evaluated by Monte Carlo.  These values provide starting benchmarks to consider together  in assessing scenario-reference concordance.  

Further, the reference sample can be regarded  as defining an importance sample (IS) for Monte Carlo integrations with respect to each of the $p_j(\y)$ as well as the mixture $\fya$ at any $\balpha.$ This recognition provides additional benchmarks from IS that can provide insights into concordance.   In general, write $h(\y)$ for a target p.d.f. in importance sampling with $p(\y)$ as proposal.  The normalised IS weights  $w^i \propto h(\y^i)/p(\y^i)$  define the discrete distribution $\{ \y^i, w^i \}_{i=\seq 1n}$  
underlying MC approximations to expectations of functions under $h(\y).$   A proviso is that $\py$ is a relevant importance sampling proposal; in particular, it should be at least as heavy-tailed as $h(\y)$.  Standard IS efficiency measures including  the empirical \% effective sample size 
$\textrm{ESS}=n^{-1}100/\sum_{i=\seq1n} (w^i)^2$ provide guidance:  initial analysis generating a relatively low ESS guides 
choice of a larger sample size.  
Applying this to cases when $h(\y)=p_j(\y)$ for mixture component $j$ defines component-specific ESS measures, and then with 
$\h(\y)=\fya$ at any chosen $\balpha$ gives ESS for the mixture.   


\section{Example: Macroeconomic Scenario Synthesis\label{sec:ascenariopps}}

\subsection{Macroeconomic Setting: Context and Perspective} 

The example comes from an area of initial motivation for the new methodology, that of scenario evaluation in macroeconomic monetary policy settings. 
Macroeconomic policy institutions such as central banks rely heavily on forecasting methods, with policymakers considering  multiple forecast summaries from research groups using structural macroeconomic models, reduced-form empirical models, and judgemental approaches. 
This typically underlies a so-called central-- or {\em baseline scenario}-- whose forecasts define a benchmark for policy path discussions.  
Overlaid on the baseline are ranges of so-called {\em alternative scenarios} based somewhat loosely on collective discussions about potential economic and financial developments; these typically lead to scenario forecasts that are location shifts of the baseline. 
In theoretical setting of this paper,  scenarios denoted by $\scn$ define forecast p.d.f.s: $\scn_0$ refers to the baseline forecast generating $p_0(\y),$ while $\scnj$ for $j=\seq 1J$ generates forecast p.d.f. $p_j(\y)$ under each of a chosen set of $J$ alternative scenarios.

Complementing this core economics-based forecasting enterprise is the parallel use of more empirical models, typically based on flexible statistical methods such as quantile regression, among others. Bayesian approaches are increasingly adopted due to the role of natural regularisation in dealing with many parameters, the ability to address coherent uncertainty quantification and the ease of incorporating  subjective inputs. In our context, this defines the statistical reference p.d.f  $p(\y)$ as the target for evaluation of scenarios and in resulting mixture syntheses. 

Our example concerns macroeconomic forecasts and policy decisions at the U.S. Federal Reserve Board (FRB). DM in our framework is effectively the Federal Open Market Committee (FOMC), whose policy discussions are informed by multiple inputs relevant to forecasting the future development of the U.S. macroeconomy under various scenarios. The illustration here focuses on December 2020 forecasts of 2021 outcomes;  $\y=y$ is univariate and represents real GDP growth; forecasts of GDP are key inputs into policy decisions of all central banks.

Baseline and scenario forecasts come from the Tealbook (TB) prepared by FRB staff for the FOMC; the TB has provided  research staff forecasts for the U.S. macroeconomy since the 1960s.   The December 2020 Tealbook reports forecasts for 2021 GDP among several macroeconomic indicators; this 
underlies $p_0(y)$. This baseline is of prime interest as it reflects the FRB staff's judgemental forecast and related uncertainty about evolving macroeconomic and financial conditions in the coming year. The alternative scenarios generate forecasts under various conditioning assumptions,   typically quantified using formal macroeconomic models. The latter can involve ranges of economic modelling choices. The scenarios are designed to assess the sensitivity of the outlook to hypothetical economic and financial developments judged sufficiently plausible to warrant consideration in policy deliberations. These forecasts underlie scenario p.d.f.s $p_j(y)$ for $j=\seq 1J.$ 

In contrast, the reference forecast distribution comes from the New York Federal Reserve (NY Fed) and is based on their statistical Outlook-at-Risk (OaR) methodology, a quite separate analysis approach related to the broader Growth-at-Risk (GAR) literature~\citep{AdrianEtal2016,Adrianetal2019,PlagborgEtal2020,Adrianetal2022}.  The NY Fed 2020 OaR forecasts of 2021 GDP growth underlie 
the reference $p(y)$. 


\subsection{Scenarios and Reference p.d.f.s} 

The example has $J=6$ for a total of 7 scenarios:  the baseline $\scn_0$,  5 contextually defined economic alternative scenarios  
$\scnj$ for $j=\seq 15$, and a final {\em\safehavennospace} scenario  $\scn_6$ detailed further below. 

The information from the December 2020 TB defines the baseline p.d.f. $p_0(y)$  for $y=\textrm{\%GDP}$ under $\scn_0$. This is 
a Student$-$t  distribution with 50 degrees of freedom (effectively a normal distribution); see Figure~\ref{fig:ScenarioEGfigs}.  Each of   5 alternative scenarios $\scnj$ has a p.d.f  of the same form but shifted in location and with modest inflation of scale.  The scenario location shifts relative to baseline are based wholly on the economic reasoning about potential, plausible economic and financial changes influencing 2021 economic growth as given in the TB. The small changes in spread   reflect increased uncertainties due to the fact that the {\em What-if?} considerations underlying location shifts are uncertain.   The scenario labels for $j=\seq 15$  in Table~\ref{tab:2020econsynth} refer to the background economic considerations underlying each of these 5 alternative scenario forecasts. Figure~\ref{fig:ScenarioEGfigs} shows the p.d.f.s of this primary set of alternative scenarios $\scnj$ for $j=\seq 15$ along with the baseline at $j=0.$     One contextually relevant point is that the p.d.f.s of  the baseline $\scn_0$ and the alternative $\scn_4$ 
are visually close to the reference, indicating that the analyses should favour one or both in the resulting mixture. 

Analysis includes a designated {\em \safehavennospace} scenario $\scn_J$ with, in this example, $J=6.$  This is designed to address potential incompleteness of the {\em core set of scenarios}-- the baseline and initial alternative scenarios $\scn_j$ for $j=\seq 0{J-1}$. The synthetic scenario $\scn_J$ has p.d.f. $p_J(y)$ that assigns mass to outcomes that are relatively poorly supported under that core set,   being over-dispersed relative to any average of them. This may then provide support for more extreme outcomes that the reference may favour.  This  addresses, in part, the reality that the specified scenario set is likely incomplete in the sense that any discrete mixture without such a  \safehaven can adequately approximate the reference. The introduction of a \safehaven component of the mixture follows theoretical developments of Bayesian predictive synthesis (BPS: \citealp[e.g.][]{TallmanWest2023,JohnsonWest2024}) that show how traditional Bayesian model uncertainty analysis and model averaging (BMA) are naturally extended to allow for {\em model set incompleteness}, i.e., the notion that all models are wrong.  Here \lq\lq models'' are replaced by \lq\lq scenarios'' to deliver this practical extension of the core, economically argued scenario set to include a \safehavennospace. Figure~\ref{fig:ScenarioEGfigs} also shows the p.d.f. $p_J(y)$ at $j=J=6$ of the \safehaven in the current example. 

The reference  p.d.f. $p(y)$ is that of a skew-T distribution based on the set of predictive quantiles reported in NY Fed's OaR analysis.  This has low degrees-of-freedom reflecting uncertainty with rather heavy tails compared to the scenarios.  
 The reference  also has some modest  negative skewness that increases p.d.f. values in the lower tail relative to the upper tail. While the heavy tails and skewness are perhaps not so evident visually in the p.d.f. plot in Figure~\ref{fig:ScenarioEGfigs} (but are clear in comparison of log p.d.f.s, not shown) these features play important roles in the evaluation of EMR and other measures in the concordance and synthesis analyses.

\subsection{Mixture Synthesis Analyses} 

\begin{figure}[p]  
    \centering
	\includegraphics[clip,width=3.2in]{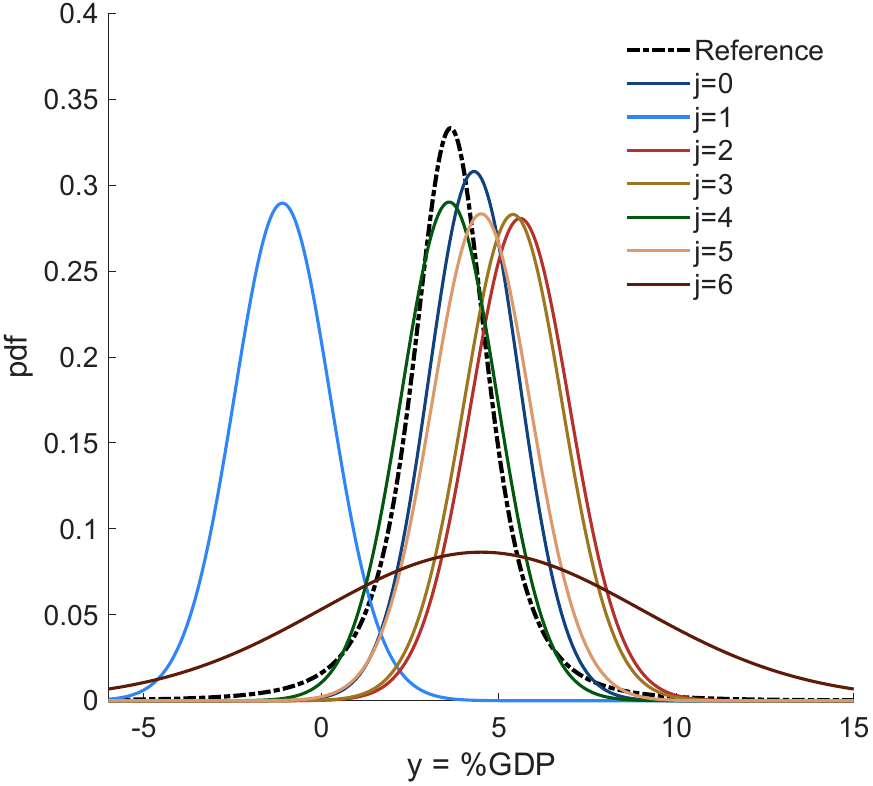}  
	\includegraphics[clip,width=3.2in]{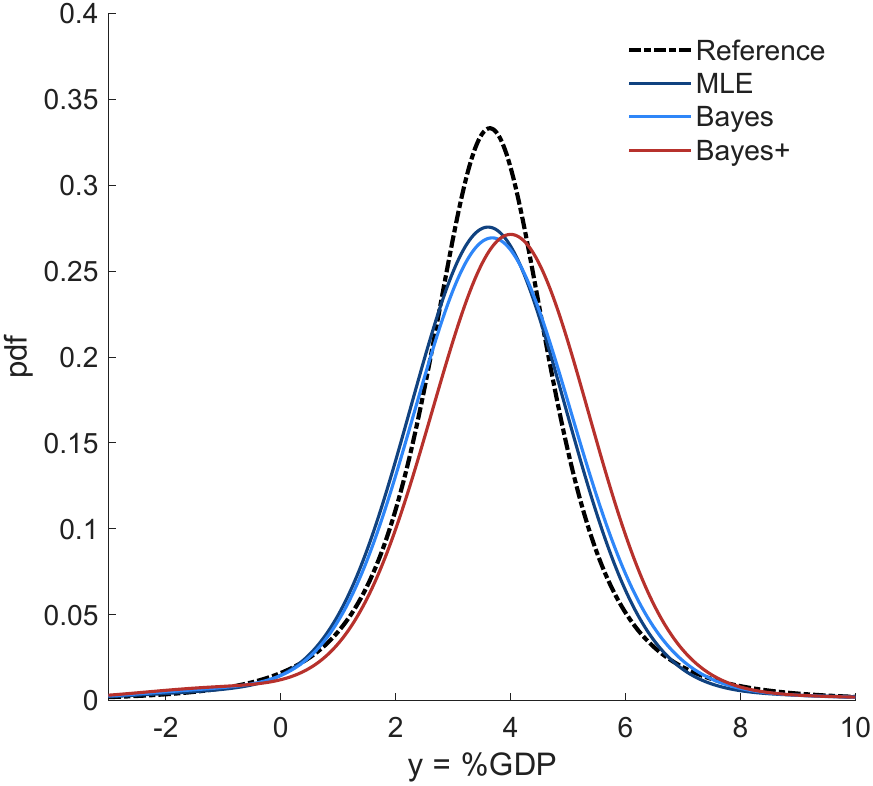}  
	\caption{Input and output p.d.f.s in the 
     2020 Macroeconomic scenario synthesis example.
     {\em\blu Left frame:} p.d.f.s of the reference and set of scenario distributions. 
                      {\em\blu Right frame:} p.d.f.s of the reference and three mixture synthesis distributions based on different $\balpha$ weight vectors, as discussed in text.                    
	 \label{fig:ScenarioEGfigs}}
\bigskip\bigskip\bigskip
	\begin{tabular}{llrrrrrrr}    
	\blu $j$ &   \blu Scenario $\scnj$&  \blu ESS\% &  \blu EMR & \blu $\widehat\alpha_j$ &\blu  $\alpha_j^{KL}$ & \blu $\alpha_j^{TV}$ &\blu  $\alpha_j^*$ &\blu  $\alpha_j^{*+}$ \medskip  \\
	\blu 0  & Baseline                                            &  74.8    &    0.462   &   0.008    &  0.005  & 0.060   &  0.047  &  0.452 \\
	\blu 1  &  2nd round severe restrictions               &   2.9    &    0.085   &   0.010    &  0.009  & 0.006   &  0.015  &  0.019 \\
	\blu 2  &  Early vaccine                                     & 24.3    &    0.323   &   0          &  0        & 0.004   &  0.012  &  0.006 \\
	\blu 3  &  Additional fiscal support                      & 29.2    &   0.347    &  0           &  0        & 0.004  &  0.013  &  0.006 \\
	\blu 4  &  Inflationary pressures                         &  94.8    &    0.490   &  0.926    &  0.926  & 0.909   &  0.838  &  0.452 \\
	\blu 5  &  Lower inflation expectations                & 64.9    &   0.449    &   0          &  0        &  0.013 &   0.028  &  0.015 \\
	\blu 6  &  Backstop                                          &  15.2    &   0.346   &  0.056     &  0.060  &  0.003  &  0.047  &  0.049 \\
	\\
	&       MLE synthesis:                                     & 97.3    &   0.497        \\
	&       Bayes synthesis:                                  & 96.7     &   0.496      \\     
	&       Bayes+ synthesis:                                & 91.1     &   0.490      \\   
	\end{tabular}
	\captionof{table}{2020 Macroeconomic scenario synthesis summaries. EMR and $\alpha$ values are rounded to 3 decimal places (full precision versions of $\alpha$ values in any column sum to 1). }\label{tab:2020econsynth} 
\end{figure}

The EMR and parameter optimisation results are summarised in Table~\ref{tab:2020econsynth}.  Analyses are based on a Monte Carlo reference random sample $(n=10^6)$.     The component-specific empirical ESS\% in the table indicate an initial ranking of scenarios relative to the reference in that measure; one detail is that $p_4(y)$ scores most highly and better than the baseline $p_0(y)$ that is {\em prima facie} favoured on economic grounds.   The corresponding
component-specific EMR values support this and give a sense of how well the scenarios are probabilistically discriminated from the reference. 

Results from three primary optimisation analyses are summarised in the table. Optimal $\balpha$ vectors are defined by maximising~\eqn{LogpostalphaDirprior} 
with three variants: 
(i) $\epsilon=0$ so defining the global EMR maximiser $\alh$ and labelled the MLE analysis; 
(ii) the default $\epsilon=0.005/(J+1)$ to induce slight shrinkage away from simplex boundaries, yielding $\als$ and labelled the Bayes analysis; 
and 
(iii) as in (ii) but subject to the additional constraint that $\alpha_0\ge \alpha_j$, $j=\seq1J,$  yielding  $\balpha^{*+}$ and labelled the Bayes+ analysis.  

The table highlights shrinkage induced by the regularisation penalty in 
comparison of  $\alh$ and  $\als.$  The raw EMR maximising solution has zeros on three mixture components and small values elsewhere, and these are moved away from the simplex boundaries under the regularised solution. 

As noted above, the baseline $p_0(y)$ has distinguished status relative to the other components, being regarded as a main, consensus forecast based on a range of inputs from economic models and discussions. Hence DM will generally have the prior expectation that $p_0(y)$ will be favoured in the mixture synthesis. However, as already reflected in the ESS measures, the optimised  $\alh$ and $\als$ very strongly weight $p_4(y)$. The scenario $\scn_4$ is most strongly concordant with the reference predictions and dominates all others including the baseline, contrary to initial expectations of the DM on contextual grounds.   The additionally constrained Bayes+ solution $\balpha^{*+}$ is relevant in this economic context, reflecting a position that DM could take to defend the baseline reasoning. Suppose the DM  {\em a priori} constrains the mixture optimisation such that the baseline will have no less weight than any other scenario; optimisation results split the weights equally between the baseline and the otherwise dominant $\scn_4$ with modest impact on the weights of other components.   For  future reference,  the synthetic posterior interpretation of the objective function in Section~\ref{sec:mixturesEMRregularised}
suggests opportunities for alternative approach to embedding such prior contextual expectations. 
The DM could consider applying soft and informed constraints by replacing the symmetric Dirichlet prior penalty with one based on an informative Dirichlet that more heavily weights some scenarios than others; in particular, this would naturally strongly favour higher values of the baseline weight $\alpha_0.$ 

The table also shows the optimal mixture weight vector based on 
minimising KL divergence of the mixture from the reference $(\balpha^{KL})$, as well as that from
minimising TV distance between the two $(\balpha^{TV})$. The former is a convex optimisation so the reported
$\balpha^{KL}$ is unique and easily found. The TV optimisation is non-convex, however; the reported $\balpha^{TV}$ is based on 1{,}000 repeat optimisations from random starting points on the simplex (that identified several local minima). 
Note the similarity of optimising weights with those from the unconstrained EMR maximisation and that TV generates slightly less extreme values.

Figure~\ref{fig:ScenarioEGfigs} displays the mixture synthesis p.d.f.s based on $\alh$ (MLE),  $\als$ (Bayes) and $\balpha^{*+}$ (Bayes+) together with the reference.  There is generally close agreement between the three mixtures though Bayes+ is a little shifted to the right. The latter feature arises since the baseline p.d.f. favours slight higher values than the reference and this mixture is constrained to ensure higher probability on the baseline.  The other main feature is that none of the mixture p.d.f.s is able to capture the leptokurtosis of the reference. They are each mixtures of seven Student-$t$ distributions with 50 degrees of freedom, while the heavy-tailed reference is a skew$-t$ with just under 3 degrees-of-freedom (and slight skewness).  Only with heavier-tailed scenario forecasts could a resulting mixture synthesis provide higher visual concordance with such a heavy-tailed and centrally peaked reference p.d.f.    Nevertheless, in terms of practical differences the EMR-based concordance of the mixtures with the reference is at or about 0.49; these values are shown in the final 3 rows of  Table~\ref{tab:2020econsynth} along with the corresponding ESS measures. At a practical level, DM would expect to find it difficult to discriminate the reference from any one of these mixture distributions based on a single observation (whether that observation is drawn from the reference or the mixture).

\section{Summary Comments} 

The foundational basis and theoretically-grounded intuition underlying EMR make it a natural, interpretable and practical concept in representing distributional concordance. 
In our context of comparing and choosing parametrised distributions based on concordance with a given reference distribution,  optimisation of EMR-- or its regularised and/or 
constrained variants-- is a natural Bayesian decision driven by outcome-based misclassification rates defining implicit bounded utilities. Intimate connections with KL divergence highlight a central position for EMR among other measures, and argue for it as a preferred measure based on natural interpretation, symmetry, boundedness, and relative insensitivity to distributional tails since it emphasises fully probabilistic concordance. 
In the optimisation problem, the dual interpretation of EMR via a purely synthetic likelihood connects to Bayesian emulation for decisions as well as providing an absolute, interpretable basis for specifying regularisation penalties via synthetic priors. 
In the specific-- but far-reaching-- setting of mixture synthesis, additional theoretical developments underlie easy and efficient implementation exploiting convex optimisation and simple, direct 
Monte Carlo simulation to evaluate the expectations underlying EMR. 

Discussion has touched on potential extensions in both theoretical and methodological areas, as well as other applications. At a theoretical level, relationships of EMR to KL divergence have potential for deeper theoretical developments. Relationships of EMR with other concordance measures may be investigated, in part, by exploiting the extensive theory of divergences.  In terms of methodology, the opportunity to embed the optimisation problem in a synthetic Bayesian emulation setting will make choice and specification of  regularisation penalty functions more directly accessible due to interpretation as smoothness priors. Integrating contextual information to define soft constraints this way will be of interest and potential importance in applications. 

On applications {\em per se}, the specific example of mixture synthesis is of central importance and of growing interest in the macroeconomic forecasting setting.
Further developments in this applied area might include analyses with higher-dimensional outcomes $\y$ (both in terms of additional variables and forecast horizons), use of the approach to help to guide the construction of new scenarios, and representation of disagreements in policy committees discussions, among other topics. Related applied opportunities exist in areas of financial risk assessment, stress testing, and business forecasting, where baseline forecasts represent current standard approaches, and 
{\em What-if?} alternatives are routinely developed and assessed to potential improve and supersede the baseline.  Beyond mixtures, the framework and approach is open to extension to aid in calibrating parametrised statistical models against a standard reference so long as the density functions involved can be evaluated and the reference distribution easily simulated.

\section*{Appendices}

\subsection*{Appendix A. \label{app:AscalemixnormalsEMRbound} Examples of Theoretical Bounds on EMR} 

Settings in which the KL-based theoretical lower bound on EMR of Section~\ref{sec:EMRaspects}
 is theoretically implied include contexts where the distribution of 
$k(\y) = \log\{\py/\fy\}$ under $\cH_p$ is continuous, unimodal and symmetric. Assuming a finite mean $m$ that must, of course, be non-negative, such a distribution is a scale mixture of normals~\citep{AndrewsMallows1974,West1987}. The p.d.f.  $g(k)$ has the form
$g(k)=\tE_v[ v^{-1}\phi\{v^{-1}(k-m)\} ]$ where $\phi(\cdot)$ is the standard normal p.d.f., $v$ is a random scale parameter 
 and $\tE_v[\cdot]$ denotes expectation with respect to its distribution.  
 
Then,  recognise that $\pi(k)=1/\{1+\exp(k)\}$ is the survival function of the standard univariate logistic distribution for real-valued $k.$   
The logistic distribution is also a normal scale mixture, so $\pi(k) = 1-\tE_u[ \Phi(u^{-1}k) ]$ where $\Phi(\cdot)$ is the standard normal c.d.f., 
$u$ is the random scale parameter and $\tE_u[\cdot]$ denotes expectation with respect to its distribution. Thus $\pi(m) = 1- \tE_u[\Phi(u^{-1}m)]$.
As a result, 
$\emr =  1- \tE_{v,u}[ \int_k\Phi(u^{-1}k)v^{-1}\phi\{v^{-1}(k-m)\} dk]$ with expectation over $v,u$ (in which, implicitly, $u\ci v)$.
Routine normal theory  yields 
$\emr = 1- \tE_{v,u}[\Phi(w^{-1}m)]$ where $w=(v^2+u^2)^{1/2}.$    
Hence $\tE_k[\pi(k)] - \pi(m) = \tE_{v,u}[\Phi(u^{-1}m) - \Phi(w^{-1}m)].$   Now, $m\ge 0$ and $w>u$ so that 
$\Phi(u^{-1}m) - \Phi(w^{-1}m) \ge 0$  implying that $\tE_k[\pi(k)] \ge \pi(m),$  as required.  This inequality is strict unless $k=v=0$.  

The analysis above may extend to more general cases when $g(k)$ is not symmetric.  Suppose, for example, that $g(k)$ is a scale mixture of skew-normal distributions~\citep{Azzalini2013}.  This is a rich class of unimodal distributions with ranges of asymmetries; it includes the above symmetric distributions as special cases.  Convolutions of skew-normals with normals are skew-normals, so it is reasonable to ask if the above development generalises. There may be broader generalisation to other classes of skewed distributions such as have been exploited in other areas~\citep[e.g.][]{DuranteEtAl2026rssSkew} so long as $g(k)$ is unimodal with  $m>0$ and/or $\tPr(k\ge 0)>0.5$. This is an aside and beyond current scope, but suggests further theoretical study.    

\subsection*{Appendix B. \label{app:BMLEconvex} Convexity and Uniqueness in EMR Mixture Optimisation} 

Referring to Section~\ref{sec:mixturesEMR}, for any $\y$ define the $(J+1)-$vector $\p(\y)=[p_0(\y),\ldots, p_J(\y)]'$. It is then easily shown that derivatives of EMR in~\eqn{EMRa} are
$$ 
  \h(\balpha) \equiv \frac{\delta\emr(\balpha)}{\delta\balpha} = \int_\y \p(\y) h(\y|\balpha) d\y
\quad \textrm{and} \quad
\H(\balpha) \equiv \frac{\delta^2\emr(\balpha)}{\delta\balpha\balpha'}
= -2\int_\y \p(\y)\p(\y)' H(\y|\balpha) d\y
$$
where $h(\y|\balpha) = \py^2/\{\fya+\py\}^2$ and $H(\y|\balpha) = h(\y|\balpha)/\{\fya+\py\}.$  
At any $\balpha$ the Hessian matrix is $\H(\balpha) = -2 \,\tE[ \p(\y)\p(\y)'  a(\y|\balpha) | \cH_p]$ where 
$a(\y|\balpha) = \py/\{\fya+\py\}^3$.
Since $a(\y|\balpha)>0$ for all $\y,\balpha$ the expectation is a positively weighted average of  rank-one 
matrices $\mathbf{p}(\y)\mathbf{p}(\y)'.$ Whether  $\y$ is continuous or discrete (the latter with at least $J+2$ support points) 
$\H(\balpha) $ is full rank and strictly negative definite for all $\balpha.$ 
Note that the discrete case is relevant when using a sample from $\py$ for MC integration to evaluate $\emr$ and its derivatives.

It follows that maximising $\emr(\balpha)$ over the simplex is a convex optimisation problem with a unique maximising value $\alh$.
Standard constrained optimisation algorithms apply.   
The extension to add priors/regularisation penalties in Section~\ref{sec:mixturesEMRregularised} maintains convexity and ensures a unique posterior mode $\als$ given any  $\epsilon>0.$  This follows since the implicit Dirichlet prior is strongly unimodal.  Again, standard constrained, non-linear optimisation methods~(e.g., the default interior-point algorithm in the {\tt fmincon} function in~\citealp{MatlabOPT}) apply and are fast and efficient.

As an aside but of some broader interest, 
the same approach shows that minimising $\klpf$ or $\klfp$ (when finite) with respect to $\balpha$ are also convex optimisations with unique solutions and similar characteristics. This also applies with any value of $\epsilon$ in the fully Bayesian version that adds the prior penalty based on a very diffuse but proper Dirichlet prior that supports non-zero $\alpha_j$ with probability one.  This links to an existing literature on sparsity and stability of KL-optimal mixtures in the context forecast combination~\citep[e.g.][]{conflitti2015optimal,DIEBOLD2023,crump2024changing,de2024multiplicative}.  Then, relative to EMR, the KL analysis typically leads to more zeros among the optimising values of $\balpha$ i.e., a sparser mixture more aggressively favouring  just one or a small number of scenarios.  This arises since KL involves expectations of the \bem{unbounded} function $\log\{\py/\fya\}$ and is very dependent on behaviour of the tails of the two p.d.fs.  This also relates to the caveat that 
KL divergence may simply be undefined in important practical contexts depending on the relative tail behaviour of $\py$ and $\fya.$ 
In contrast, EMR is more conservative (and numerically more robust) in discounting scenarios that are less concordant with the reference though not extremely so; this arises as EMR is based on expectations under $\cH_p$ of the bounded function $\fya/\{ \fya+\py\}$  That said, the full shrinkage to boundaries of the simplex still arises and can be ameliorated, if desired, by modest regularisation as provided by the penalty induced under the minimally informative synthetic prior in Section~\ref{sec:mixturesEMRregularised}.

\subsection*{Appendix C. Regularisation Parameter Specification\label{sec:calibrateepsilon} } 

In the setting of Section~\ref{sec:mixturesEMRregularised}, DM aims
 to calibrate the choice of small $\epsilon>0$ and reasons by analogy with the canonical setting for Dirichlet prior/posterior distributions, i.e., 
 that of multinomial sampling. With $\balpha$ representing scenario probabilities, the least informative multinomial sample is just one draw from one of the scenarios; the implied Dirichlet posterior 
then has parameter updated by $+1$ in one element only. 
With no loss of generality, suppose a single outcome is known to come from the baseline; this single draw posterior is then $\tDir(\bone(1+\epsilon)+\e)$ where $\e=(1,0,\ldots,0)'$.  Now, 
to reflect a minimally informative setting, suppose the posterior is modified to $\tDir(\bone(1+\epsilon)+x\e)$ 
for some very small, positive $x$. This can be regarded as the posterior under an imaginary fractional observation; for example, $x=0.01$ says the information content of the posterior relative to the prior is 1\% of that arising on observing a single multinomial draw. 

Under this posterior with specified $x,$ the prior mode $1/(J+1)$ increases to posterior mode $\alpha_0^*=(\epsilon+x)/\{(J+1)\epsilon+x\}$ on $\scn_0$, and decreases to  $\alpha_j^*=\epsilon/\{(J+1)\epsilon+x\}$ on the other scenarios $j>0.$ In this minimal information context it is rationale to limit this latter \lq\lq shrinkage towards zero; suppose DM reflects this by requiring that $\alpha_j^*\ge p/(J+1)$ for some fractional reduction $p\in (0,1).$  This implies $\epsilon\ge cx/(J+1)$ where
$c=p/(1-p)$.  Here $c$ is explicitly a lower bound on the reduction from prior to posterior {\em odds} on $\scnj$ for $j>0$ given the minimal information of a single outcome under $\scn_0.$  For example, the choice $c=0.5$ limits this odds reduction to no more than 50\%.   The choices $x=0.01$ and $c=0.5$  imply $\epsilon\ge 0.005/(J+1),$ and this value is recommended as a default.


\subsection*{Conflicts of Interest and Funding}
The authors have no conflicts of interest nor any funding sources to report. 
   
\subsection*{Code and Data}
The authors provide short, commented code and input data to reproduce the results in the example of Section 5.

\setstretch{1.0} 
\bibliography{EMRetal2026}
\bibliographystyle{chicago}

\end{document}